\documentclass[12pt,a4paper]{article}
\usepackage{color,graphicx}

\def\lvec#1{\setbox0=\hbox{$#1$}
    \setbox1=\hbox{$\scriptstyle\leftarrow$}
    #1\kern-\wd0\smash{
    \raise\ht0\hbox{$\raise1pt\hbox{$\scriptstyle\leftarrow$}$}}
    \kern-\wd1\kern\wd0}
\def\rvec#1{\setbox0=\hbox{$#1$}
    \setbox1=\hbox{$\scriptstyle\rightarrow$}
    #1\kern-\wd0\smash{
    \raise\ht0\hbox{$\raise1pt\hbox{$\scriptstyle\rightarrow$}$}}
    \kern-\wd1\kern\wd0}

\def\diracstar#1#2{
    \setbox0=\hbox{$\gamma$}\setbox1=\hbox{$\gamma_{#1}$}
    \gamma_{#1}\kern-\wd1\kern\wd0
    \smash{\raise4.5pt\hbox{$\scriptstyle#2$}}}

\newcommand{\beq}{\begin{equation}}
\newcommand{\eeq}{\end{equation}}
\newcommand{\beqn}{\begin{eqnarray}}
\newcommand{\eeqn}{\end{eqnarray}}
\newcommand{\nn}{\nonumber}

\begin{document}
\begin{titlepage}
\pagestyle{empty}
\date{}
\title{
\bf Note on lattice regularization and equal-time correlators for parton distribution functions
\vspace*{3mm}}

\author{
G.C.\ Rossi$^{a)\,b)}$  \, M.\ Testa$^{c)}$
}
\maketitle
\begin{center}
  $^{a)}${\small Dipartimento di Fisica, Universit\`a di  Roma
  ``{\it Tor Vergata}'' \\ 
  INFN, Sezione di Roma 2}\\
  {\small Via della Ricerca Scientifica - 00133 Roma, Italy}\\
  \vspace{.2cm}
  $^{b)}${\small Centro Fermi - Museo Storico della Fisica e Centro Studi e Ricerche ``E.\ Fermi''}\\
  {\small Piazza del Viminale 1 - 00184 Roma, Italy}\\
    \vspace{.2cm}
    $^{c)}${\small Dipartimento di Fisica, Universit\`a di  Roma
  ``{\it La Sapienza}'' \\ 
  INFN, Sezione di Roma  ``{\it La Sapienza}''}\\
    {\small Piazzale A.\ Moro 5 - 00187 Roma, Italy}\\
\end{center}

\abstract{We show that a recent interesting idea to circumvent the difficulties with the continuation of parton distribution functions to the Euclidean region, that consists in looking at equal time correlators between proton states of infinite momentum, encounters some problems related to the power divergent mixing pattern of DIS operators, when implemented within the lattice regularization.
}
\end{titlepage}
\newpage
     
\section{Introduction}
\label{sec:INTRO}

It would be of the utmost phenomenological importance to be able to compute the parton distribution function directly from first principles in lattice QCD (LQCD) rather than reconstructing it from its moments. In the most direct naive approach this program is obstructed by the impossibility of performing the Wick rotation which would allow to express the Minkowski amplitude in terms of Euclidean quantities, suitable for LQCD simulations. 

To bypass this difficulty it has been suggested in ref.~\cite{Ji:2013dva} to work with the equal-time (E-T) product of two currents taken between proton states in the limit of infinite three-momentum. This quantity can be directly computed in Euclidean region. Formally, i.e.\ ignoring renormalization effects, this procedure yields the correct Bjorken limit for the imaginary part of the matrix element of the product of two currents taken close to the light-cone (L-C) between proton states at rest.

We show in the present note that the interesting proposal of ref.~\cite{Ji:2013dva} is however still insufficient to implement this program on the lattice, because of the need of power divergent subtractions required to renormalize short-distance DIS operators of {\it any} dimension. These divergences are due to the existence of trace operator mixings, formally irrelevant on the light-cone (in Minkowski space $x^2=0$), but affecting the construction of renormalizable DIS operators at space-like separation (in Euclidean space, $x^2=-z^2\neq 0$).

The plan of the paper is as follows. In sect.~\ref{sec:PEC} we illustrate the problem with the continuation to the Euclidean region of the amplitude whose imaginary part yields, in the Bjorken limit, the DIS cross section. In sect.~\ref{sec:SFETC} we review the strategy proposed by Ji in his seminal paper~\cite{Ji:2013dva} to formally circumvent this difficulty, and we illustrate the difficulties posed by the needs of renormalization in the calculation of E-T correlators, which apparently prevent the naive practical lattice implementation of the proposal. Short conclusions can be found in sect.~\ref{sec:CONC}. In Appendix~A we show that formally (i.e.\ ignoring renormalization effects) the proposal of ref.~\cite{Ji:2013dva} indeed leads to the standard expression of the DIS structure function. In Appendix~B we illustrate in a simple toy-model in which way structure functions are deformed if power divergent mixings are not properly taken care of. 

\section{The problem with the Euclidean continuation}
\label{sec:PEC}

\subsection{Generalities}
\label{sec:GEN}

In this section we want to illustrate the nature of the problem one encounters with the Euclidean continuation of the hadronic matrix elements of the product of two currents.

To reduce the argument to its essentials and avoid irrelevant (for the purpose of this paper) kinematical complications we drop all flavour and Lorentz indices on the hadronic currents. We shall then consider a hypothetical theory of ``scalar quarks'' in which an appropriately renormalized scalar current, $j(x)=\phi^2(x)$, carrying momentum $q$ ($q^2 < 0$) hits a scalar ``proton'' at rest. The inclusive cross section of this process is proportional to
\begin{eqnarray}
W(q^2,q\cdot p) \equiv \int d^4 x \, e^{i q x} \langle p| j(x) j(0) |p\rangle \, . \label{amplitude2}
\end{eqnarray}
In DIS experiments one is interested in the behaviour of $W$ in the Bjorken limit 
\begin{eqnarray}
q^2 \rightarrow -\infty
\end{eqnarray}
with the ratio 
\begin{eqnarray}
\omega \equiv - \frac {q^2}{2p\cdot q} 
\label{omega}
\end{eqnarray}
fixed. The spectral condition gives for $\omega$ the allowed kinematical region
\begin{eqnarray}
0 \leq \omega \leq 1 \, . \label{allowed}
\end{eqnarray}
In the Bjorken limit eq.~(\ref{amplitude2}) is dominated by the light-cone singularities of the product of two currents giving rise to an asymptotic expansion of the form
\begin{eqnarray}
\langle p | j(x) j(0) |p\rangle \stackrel{x^2 \approx 0} {\approx}\Delta (x^2) \sum_{n=0}^{\infty} \alpha_n (\mu^2 x^2) x^{\mu_1} \dots x^{\mu_n} \langle p |\tilde O^{(n)}_{\mu_1 \dots \mu_n} (0) |p\rangle \, , \label{ligopexp}
\end{eqnarray}
where $\Delta(x^2)$ is the free scalar propagator (see eq.~(\ref{DELTA}) of Appendix~A), the $\alpha_n (\mu^2x^2)$ are logarithmically singular functions computable in perturbation theory and $\tilde O^{(n)}_{\mu_1 \dots \mu_n} (0)$ is an appropriately renormalized version of the bare DIS operator 
\beq
O^{(n)}_{\mu_1 \dots \mu_n} (0)=\phi (0) \partial_{\mu_1} \dots \partial_{\mu_n} \phi(0)\label{OBARE}
\eeq 
with subtraction point $\mu$. It is important to keep in mind that the matrix elements of the $\tilde O^{(n)}$'s give rise to several (u.v.\ finite) form factors with tensor structures of the form
\begin{eqnarray}
&&\langle p |\tilde O^{(n)}_{\mu_1 \dots \mu_n} (0) |p\rangle = A^{(n)}(\mu) p_{\mu_1} \dots  p_{\mu_n} + B^{(n)}(\mu) p_{\mu_1} \dots g_{\mu_i \mu_j} \dots p_{\mu_n} +\nonumber\\
&&+ C^{(n)}(\mu) p_{\mu_1} \dots g_{\mu_i \mu_j} \dots g_{\mu_\ell\mu_k} \dots p_{\mu_n} +\dots\, , \label{traceterms}
\end{eqnarray}
with possible multiple insertions of the metric tensor. Such terms are subdominant in the light-cone expansion~(\ref{ligopexp}) with respect to the first one, $A^{(n)}$, and can be  consequently dropped. Therefore in the Bjorken limit we get (see Appendix~A) 
\begin{eqnarray}
&&W (q^2, q\cdot p) \approx  \nonumber \\
&&\approx\int d^4 x \, e^{i q x}  \Delta (x^2) \sum_{n=0}^{\infty} \alpha_n (\mu^2 x^2) A^{(n)} (\mu)(p \cdot x)^n  \approx \frac {\omega f (\omega, q^2)}{-q^2} \, .\label{wx}
\end{eqnarray}
Eq.~(\ref{wx}) yields the structure function in terms of the finite matrix elements, $A^{(n)}(\mu)$, defined in eq.~(\ref{traceterms}), in the resummed form~\footnote{The precise relation between the coefficients $\alpha_n (\mu^2 x^2)$ in $x$-space and  $\beta_n (q^2/\mu^2)$ in Fourier space is worked out in the classical book of ref.~\cite{Yndurain}, and it is of no interest in this discussion.} 
\begin{eqnarray}
f (\omega, q^2) = \sum_{n=0}^{\infty} (-1)^n \beta_n (q^2 / \mu^2) A^{(n)}(\mu) \delta^{(n)} (\omega) \, , \label{momentsform}
\end{eqnarray}
where $\delta^{(n)}$ is the $n$-th derivative of the Dirac $\delta$-function (see eq.~(\ref{DELTN})). Eq.~(\ref{momentsform}) provides a formal definition of the structure function $f(\omega, q^2)$. The absolute normalization of the $\alpha_n$ coefficients in eqs.~(\ref{ligopexp}) and~(\ref{wx}) is fixed by matching with the expansion of $\langle p | j(x) j(0) |p\rangle$ in perturbation theory. 

An important property, stemming from eqs.~(\ref{amplitude2}) and~(\ref{momentsform}), is that the support of $f(\omega, q^2)$ in the variable $\omega$ is given by eq.~(\ref{allowed}). Eq.~(\ref{momentsform}), together with crossing symmetry, also implies the well known relation between the matrix elements of the local operators $\tilde O^{(n)}$ and the moments of the structure functions~\footnote{The integral over $\omega$ should in fact be extended from $-1$ to $+1$, but crossing symmetry allows the restriction to the $[0,1]$ interval.} expressed by the relations 
\begin{eqnarray}
\int_0^1 d \omega f (\omega, q^2) \, \omega^n \approx \beta_n (q^2 / \mu^2) A^{(n)} (\mu) \, . \label{singlemom}
\end{eqnarray}
Eq.~(\ref{singlemom}) has been used several times in order to get non-perturbative informations on structure function moments from LQCD. 

It would clearly be of a great interest to find a resummation of eq.~(\ref{momentsform}) allowing the direct computation of the structure function, starting from the Euclidean lattice regularized QCD. 

\subsection{Euclidean continuation}
\label{sec:EC}

The most direct way of determining $f(\omega, q^2)$ would be to compute it in lattice simulations starting from eq.~(\ref{amplitude2}). Unfortunately it is not possible to perform the Wick rotation to continue the Minkowski amplitude into the Euclidean region suitable for lattice QCD simulations. Let us in fact consider the Minkowski amplitude
\begin{eqnarray}
T(q,p) \equiv \int d^4 x \, e^{i q x} \langle p| T (j(x) j(0)) |p\rangle \, , \label{amplitude}
\end{eqnarray}
the imaginary part of which is $W(q^2,q\cdot p)$. If in eq.~(\ref{amplitude}) we perform the {\em change of variables}
\begin{eqnarray}
x^0 &=& - i x^0_E \label{CHAN1} \\
{\bf r} &=& {\bf r}_E\, ,\label{CHAN}
\end{eqnarray}
we can express the Minkowski amplitude in terms of Euclidean quantities in the form 
\begin{eqnarray}
T(q, p) =-i \int d^4 x_E \, e^{q^0 x^0_E} \langle p| T_E (j(x_E) j(0)) |p\rangle  e^{-i {\bf q}\cdot {\bf r}} \, .\label{amplitude1}
\end{eqnarray}
Eq.~(\ref{amplitude1}) is a meaningful formula under the condition that it is well defined. Due to the presence of the growing exponential $e^{q^0 x^0_E}$ we must worry about the behaviour of the Euclidean $T$-product as $x^0_E \rightarrow +\infty$. 

With the definition 
\begin{eqnarray}
F(x^0_E) \equiv \int d {\bf r} \langle p| T_E (j(x_E) j(0)) |p\rangle e^{-i {\bf q}\cdot {\bf r}} \, ,
\end{eqnarray}
we have
\begin{eqnarray}
F(x^0_E) \stackrel {x^0_E \rightarrow + \infty} {\rightarrow}(2\pi)^3 \sum_n |\langle n|j(0)|p\rangle|^2 e^{-(E_n-m) x^0_E} \delta ({\bf p}_n-{\bf q}) \, ,
\end{eqnarray}
so that the condition under which the change of variables~(\ref{CHAN1})-(\ref{CHAN}) is meaningful, is
\begin{eqnarray}
E_n-m > q^0 \, .\label{DIS}
\end{eqnarray}
On the other hand $T(q,p)$ in eq.~(\ref{amplitude}) develops an imaginary part, $W(q,p)$, only if an on-shell intermediate state can be created, i.e.\ only if 
\begin{eqnarray}
E_n-m = q^0 \, ,
\end{eqnarray}
so, looking at eq.~(\ref{DIS}), we conclude that while working in the Euclidean region one cannot access $W(q,p)$.

In view of this obstruction some new strategies have been tried. A particularly promising one is the approach proposed in~\cite{Ji:2013dva} and developed in refs.~\cite{Xiong:2013bka,Ma:2014jga,Lin:2014zya,Ji:2015jwa,Alexandrou:2015rja,Zhang:2017bzy,Alexandrou:2017huk,Orginos:2017kos}, which we shall now discuss.

\section{Structure functions from equal-time correlators}
\label{sec:SFETC}
 
In our scalar model the proposal made in~\cite{Ji:2013dva} amounts to compute the (bare) structure function from the formula 
\begin{eqnarray}
F (\omega) = \lim_{P_z \rightarrow +\infty} \frac {P_z}{2 \pi} \int_{-\infty}^{+\infty} d z\, e^{iz \omega P_z} \langle P_z|\phi (0) \phi(z)|P_z\rangle\, , \label{proposal}
\end{eqnarray}
where $|P_z\rangle$ denotes the state of a proton with momentum $P_z$ along the $z$-axis and $z$ is the space-time event $(0,0,0,z)$. Eq.~(\ref{proposal}) expresses $F(\omega)$ in terms of the matrix element of a $x_0=0$ operator which thus takes the same value in Minkowski as well as in Euclidean time. Its computation can be thus performed in principle in lattice QCD simulations.
 
In order to see where the problems with renormalization (and in particular with power divergent operator mixings) lie, it is convenient to first rewrite eq.~(\ref{proposal}) shifting the Lorentz transformation from the proton state to the space-time argument of the bilocal operator. 

The Lorentz transformation which brings a proton with momentum $P_z$ at rest is \begin{eqnarray}
&&{x^0}' = \frac {x^0 + \beta z}{\sqrt{1-\beta^2}}\, , \qquad z' =  \frac {z + \beta x^0}{\sqrt{1-\beta^2}} \, , \label{LOR}
\end{eqnarray}
with
\begin{eqnarray}
\beta = \frac {P_z} {\sqrt {m^2 +P_z^2}} \, , \label{BLOR}
\end{eqnarray}
so that the bilocal operator matrix element can be written as
\begin{eqnarray}
\langle P_z|\phi (0) \phi(z)|P_z\rangle = \langle m| \phi (0) \phi (\frac {P_z} {m} z, 0, 0, \frac {\sqrt {m^2 +P_z^2}}{m} z )|m \rangle  \, ,\label{REST}
\end{eqnarray}
with $|m\rangle$ a proton state at rest. Inserting eq.~(\ref{REST}) in~(\ref{proposal}), one gets
\begin{eqnarray}
&&F (\omega) =\nonumber \\
&&= \lim_{P_z \rightarrow +\infty} \frac {P_z}{2 \pi} \int_{-\infty}^{+\infty} d z e^{iz \omega P_z} \langle m | \phi (0) \phi (\frac {P_z}{m} z,0,0,\frac {\sqrt{P_z^2+m^2}}{m}z) |m \rangle= \nonumber \\
&&= \lim_{P_z \rightarrow +\infty} \frac {1}{2 \pi} \int_{-\infty}^{+\infty} \, d y e^{iy \omega} \langle m| \phi (0) \phi (\frac {y}{m} ,0,0, \frac {\sqrt{1+ \frac {m^2} {P_z^2}}} {m} y) |m\rangle\, . \label{duestruc} 
\end{eqnarray}
The problem of eq.~(\ref{duestruc}) with the mixing of trace operators is best exhibited by considering its moments. Let us, for instance, compute the second moment~\footnote{The $\omega$ integration has been formally extended over the whole real axis. The support of $F(\omega)$ will take care of limiting it to the allowed region~(\ref{allowed}).}
\begin{eqnarray}
&&\int_{- \infty}^{+ \infty} \omega^2 F (\omega) d \omega = \label{SECMOM} \\
&&= \lim_{P_z \rightarrow +\infty} \frac {P_z}{2 \pi} \int_{-\infty}^{+\infty}  d \omega d z \,\omega^2 e^{iz \omega P_z} \langle m | \phi (0)  \phi (\frac {P_z}{m} z,0,0,\frac {\sqrt{P_z^2+m^2}}{m}z) |m \rangle =\nonumber \\
&&= - \lim_{P_z \rightarrow +\infty} \frac {1}{P_z^2} \int_{-\infty}^{+\infty} d z \frac {d^2 \delta (z)}{d z^2} \langle m |\phi (0) \phi (\frac {P_z}{m} z,0,0,\frac {\sqrt{P_z^2+m^2}}{m}z) |m \rangle = \nonumber \\
&&=- \lim_{P_z \rightarrow +\infty} \frac {1}{P_z^2} \frac {d^2}{d z^2} \langle m |\phi (0) \phi (\frac {P_z}{m} z,0,0,\frac {\sqrt{P_z^2+m^2}}{m}z)  |m \rangle \Big{|}_{z=0} \, . \nonumber
\end{eqnarray}
The connection between the second moment and the local operator of rank two in the E-T approach is therefore
\begin{eqnarray}
&&\int_{- \infty}^{+ \infty} \omega^2 F (\omega) d \omega = \label{moment} \\
&&=- \lim_{P_z \rightarrow +\infty} \frac {1}{P_z^2} \Big{(}\frac {P_z^2} {m^2}  \langle m |O^{(2)}_{00}(0) |m \rangle +  \frac {P_z^2+ m^2} {m^2}  \langle m |O^{(2)}_{33}(0) |m \rangle + \nonumber \\ 
&&+2 \frac {P_z \sqrt{P_z^2+m^2}} {m^2}  \langle m |O^{(2)}_{03}(0) |m \rangle\Big{)}\nonumber \, ,
\end{eqnarray}
where formally
\begin{eqnarray}
O^{(2)}_{\mu \nu} = \phi (0) \partial_\mu \partial_\nu \phi(0) \, .\label{BILOC}
\end{eqnarray}
Ignoring divergences formally everything works fine~\cite{Brandt:1972nw,Fritzsch:2003fg}. In particular we have (with $g_{\mu\nu}$ the Minkowski tensor)
\begin{eqnarray}
\langle p | O^{(2)}_{\mu \nu} |p \rangle = A^{(2)} p_\mu p_\nu + B^{(2)} g_{\mu \nu} \, ,
\end{eqnarray}
so that
\begin{eqnarray}
&&\int_{- \infty}^{+ \infty} \omega^2 F (\omega) d \omega = \label{momentcan} \\
&&= -  \lim_{P_z \rightarrow +\infty} (A^{(2)} - \frac {B^{(2)}}{P_z^2}) = - A^{(2)} \nonumber \, .
\end{eqnarray}
We now discuss what happens (within perturbation theory) {\it in the case of a renormalizable field theory}, like QCD. We will compare the case of dimensional regularization with the case of the lattice regularization. 

\subsection{Dimensional regularization}
\label{sec:DR}

Adopting dimensional regularization we will be insensitive to power divergent mixings. We must therefore only worry about the multiplicative renormalization of the bare DIS operators. 

In other words, in constructing the moment generating function it is enough to insert for every bare DIS operator $\phi (0) \partial_{\mu_1} \dots \partial_{\mu_n} \phi (0)$ the combination (see eqs.~(\ref{OBARE}) and~(\ref{momentsform}))
\begin{eqnarray}
\beta_n (q^2 / \mu^2)\, {\tilde O}^{(n)}_{\mu_1\dots \mu_n} = \beta_n (q^2 / \mu^2) \, Z_n (\epsilon, {\mu}) \, \Big{[} \phi (0) \partial_{\mu_1} \dots \partial_{\mu_n} \phi (0)\Big{]}_{\epsilon}\, , \label{OTIL}
\end{eqnarray}
where in dimensional regularization $\epsilon=4-D$. By construction the matrix elements, $A^{(n)}(\mu) $, of the operator $Z_n (\epsilon, {\mu}) \, [ \phi (0) \partial_{\mu_1} \dots \partial_{\mu_n} \phi (0) ]_{\epsilon}$ are u.v.\ finite as $\epsilon \to 0$. When multiplied by the Wilson coefficients $\beta_n (q^2 / \mu^2)$, they yield  renormalization group invariant, $\mu$-independent quantities.

In order to proceed with the construction of the properly renormalized, u.v.\ finite  structure function one introduces the analytic continuation of the quantities 
\begin{eqnarray}
&&  \beta_n (q^2 / \mu^2)  \to B (n;q^2 / \mu^2) \, ,\label{HN}\\
&& A^{(n)}(\mu) \to A(n;\mu) \label{AN}
\end{eqnarray}
to complex values of $n$. In terms of the inverse Mellin transforms
\begin{eqnarray}
&&{\cal M}_B(\omega; q^2 / \mu^2) = \frac {1} {2 \pi i} \int_L (\omega)^{-n-1} B(n; q^2 / \mu^2) \,dn \, ,\label{G}\\
&& {\cal M}_{A}(\omega;\mu)=  \frac{1}{2\pi i} \int_L \, (\omega)^{-n-1}A(n;\mu) \,d n \, ,\label{FLAMBDA}
\end{eqnarray}
the required renormalized structure function is finally given by the convolution formula~\cite{Yndurain}
\begin{eqnarray}
&&{\tilde F} (\omega,q^2) = \int_\omega^1 \frac {d \omega'} {\omega'} \, {\cal M}_B (\omega'; q^2 / \mu^2)\,{\cal M}_{A}(\omega / \omega';\mu) =\nn\\
&&=\frac{1}{2\pi i}\int_L d n \,\omega^{-n-1} B (n; q^2 / \mu^2) A(n;\mu) \, , \label{FLAMBDA1}
\end{eqnarray}
where $L$ is the line $n_0 + i \nu$ in the complex $n$ plane with $n_0$ sufficiently large to ensure convergence of the integrals. 

The moments of ${\tilde F}(\omega,q^2) $ are the matrix elements of the operators~(\ref{OTIL}) that in the limit $\epsilon \to 0$ yield finite, $\mu$-independent quantities. 

The second equality in eq.~(\ref{FLAMBDA1}) shows that the proper way to carry out the summation over moments, formally given by eq.~(\ref{momentsform}), is to perform the integral along the line $L$ in the complex $n$-plane of the analytic continuation of the $A^{(n)}(\mu)$ amplitudes (which represent the hadron matrix elements of the renormalized DIS operators) times the Wilson coefficients $\beta_n (q^2/ \mu^2)$ (which inject the information about the anomalous dimensions of the renormalized DIS operators). Clearly for the whole procedure to be meaningful, i.e.\ to yield an u.v.\ finite ${\tilde F} (\omega,q^2)$, the u.v.\ finiteness of the (physically measurable~\footnote{\,For a recent compilation of DIS data see the references listed in~\cite{Christy:2012tk}.}) moments is necessary. In the next section we show that this cannot be the case in the lattice regularization. 

\subsection{Lattice regularization}
\label{sec:LR}

In the case of lattice regularized QCD the situation is not so simple due to the appearance of two related problems. The first is the need to perform power divergent subtractions to make the $O^{(n)}$ lattice operators finite. In fact, in contrast with the usual L-C approach, trace terms in the E-T approach are not suppressed since $x^2=-z^2\neq0$. The second problem is that the support condition~(\ref{allowed}) is only guaranteed for the leading contribution $A^{(n)}$ and will be violated if trace terms are not appropriately subtracted. 

Power divergences in the cutoff ($\Lambda=a^{-1}$) appear in the moments of $F(\omega)$ of eq.~(\ref{proposal}) due, as we said, to mixing of high dimension operators with lower dimensional ones~\cite{Capitani:1994qn,Beccarini:1995iv}, preventing the $P_z \rightarrow \infty$ limit to be taken.  

In fact, referring again, as an example, to the second moment associated to the local operator~(\ref{BILOC}), we see that the contribution from the mixing of a typical lower dimensional ``trace'' operator, $a^{-2}\phi(0)^2 g_{\mu\nu}$ to eq.~(\ref{SECMOM}) is 
\begin{eqnarray}
\hspace{-1.cm}&&\int_{- \infty}^{+ \infty} \omega^2 F (\omega) d \omega \Big{|}_{\rm trace\,operator}\propto -\frac {1}{a^2 P_z^2}  \Big{(}\frac {P_z^2} {m^2} -  \frac {P_z^2+ m^2} {m^2}  \Big{)} =  \frac {1}{a^2P_z^2} \, . \label{DIVE}
\end{eqnarray}
The correct procedure would be to send $a\to 0$ first and then $P_z\to\infty$, as on the lattice the largest attainable momentum is O($a^{-1}$). So unless we perform a non perturbative subtraction of power divergent terms, the $P_z\to\infty$ limit cannot be taken.

As we recalled above, the existence of this difficulty is also signalled by a problem with the support of $F(\omega,q^2)$. In fact, the support condition eq.~(\ref{allowed}) is guaranteed by eq.~(\ref{ligopexp}) for the leading light-cone singularity. On the contrary the trace terms are not related to the current-hadron scattering and therefore will give contributions for all values of $\omega$. Their subtraction is essential for the success of the program. 

Another way to see expose these difficulties is to notice that, although the matrix element of the bilocal operator in~(\ref{proposal}) is ``well-behaved'' in $z$ (it is only logarithmically divergent for small $z$ and exponentially damped for large $z$), this circumstance is not enough to allow interpreting its Fourier transform, $F(\omega)$, as the desired parton distribution function. The reason is that the Fourier transform of a logarithm~\footnote{We recall the formula $\int e^{i z \omega} \log |z| dz = - 1/2|\omega|$~\cite{LIGHTHILL}} is a function of $\omega$, the moments of which are all divergent, unless the support of $F(\omega)$ is limited to the physical region $[-1,+1]$ which is clearly not the case in the case at hand. 

In Appendix~B we illustrate in a simple toy-model in which way structure functions are deformed if power divergent mixings are not properly taken care of.

It is important to stress that these power divergences have nothing to do with the exponentiated linear divergence related to the presence of the Wilson line which in QCD makes the bilocal operator gauge invariant. This linear divergence is not a lattice artefact. It would be there also in the continuum and it is due to the fact that the Wilson line is a non-local operator joining the points 0 and $z$~\cite{Boucaud:1989ga,Boucaud:1992nf,Martinelli:1995vj}. References~\cite{Ishikawa:2016znu,Ishikawa:2017jtf} and~\cite{Chen:2016fxx} only consider this linear divergence  and propose a method to take care of it. 

\section{Conclusions}
\label{sec:CONC}

In the limit of large proton momentum it is possible to express the Minkowski DIS structure functions in terms of E-T Euclidean correlators, as suggested in refs.~\cite{Ji:2013dva} and elaborated in~\cite{Xiong:2013bka,Ma:2014jga,Lin:2014zya,Ji:2015jwa,Alexandrou:2015rja,Zhang:2017bzy,Alexandrou:2017huk,Orginos:2017kos}.

On the lattice, however, the presence of power divergent mixings with trace operators makes the situation problematic. Such power divergences are not easy to eliminate and hinder the reconstruction of the full parton distribution function in terms of the Mellin convolution between the E-T matrix elements of renormalized, subtracted local operators and the corresponding Wilson coefficients. 

Taking as an example the second moment, we have shown that in a nutshell the problem is related to the fact that, while in Minkowski metric trace operator contributions are proportional to $a^{-2}x^\mu x_\mu=0$ (namely to a quantity which is zero on the light-cone), in the E-T approach they are proportional to the non-vanishing combination $a^{-2}z^2\neq 0$ and in matrix elements leave behind terms like~(\ref{DIVE}). 

In the absence of an appropriate non perturbative renormalization procedure trace terms will contaminate eq.~(\ref{proposal}) in an unpredictable way.

\vspace{.3cm}
\noindent {\bf Acknowledgments -} We wish to thank C.\ Alexandrou, K.\ Cichy, M.\ Constantinou, K.\ Jansen, F.\ Steffens and C.\ Wiese for discussions and their interest in this work.

\appendix 
\renewcommand{\thesection}{Appendix A} 
\section{Partons and bilocal operators}
\label{sec:APPA} 

We want to show that, ignoring renormalization effects, eq.~(\ref{proposal}) provides the correct definition of the DIS structure function~\cite{Brandt:1972nw,Fritzsch:2003fg}.

Let us consider the hadronic expression of the deep inelastic cross section in the parton approximation in the case of the scalar current $j(x) = \phi^2 (x)$. After contracting two of the $\phi$'s into the scalar propagator, $\Delta$, one gets 
\begin{eqnarray}
&&(2 \pi)^4 W (q^2, q\cdot p) = \sum_n \int \frac {d {\bf k}} {2 |{\bf k}|}|\langle n|\phi(0)|p\rangle|^2 (2 \pi)^4 \delta^4 (p+q-p_n -k) = \nonumber \\
&&= \int d^4 x e^{- i q \cdot x} \langle p|\phi (0) \phi(x)|p\rangle \Delta (x^2) \, , \label{deepin}
\end{eqnarray}
where we have introduced 
\begin{eqnarray}
&&\Delta (x^2) \equiv \int \frac {d {\bf k}} {2 |{\bf k}|} e^{i k \cdot x} = \int d^4 k \, \delta (k^2) \theta (k^0) e^{ikx} \label{DELTA}
\end{eqnarray}
with $k^\mu \equiv (|{\bf k}|, {\bf k} )$ the massless parton final momentum. 

In the {\it canonical} parton model the quantity $\langle p|\phi (0) \phi(x)|p\rangle$ is regular and is evaluated in the limit $x^2 \to 0$. So we may write
\begin{eqnarray}
&&{\tilde f} (p\cdot x) \equiv \langle p|\phi (0) \phi(x)|p\rangle \Big{|}_{x^2=0} = \int_{- \infty}^{+ \infty} d \lambda f (\lambda) e^{- i \lambda p\cdot x} \label{bilocstru1} \\
&&f (\lambda) = \frac {1}{2 \pi} \int_{- \infty}^{+ \infty}{\tilde f} (p\cdot x) e^{i \lambda p\cdot x} d (p\cdot x) \label{bilocstru2}
\end{eqnarray}
and
\begin{eqnarray}
&&(2 \pi)^4 W (q^2, q\cdot p) = \int_{- \infty}^{+ \infty} d \lambda f (\lambda) \int d^4 x \, e^{- i (q + \lambda p) \cdot x} \Delta (x^2) = \nonumber \\
&&= (2 \pi)^4 \int_{- \infty}^{+ \infty} d \lambda f (\lambda) \delta [(q + \lambda p)^2] \theta [ (q + \lambda p)_0]
\end{eqnarray}
which leads to (recall the definition~(\ref{omega}))
\begin{eqnarray}
W (q^2, q\cdot p) \approx \frac {\omega f (\omega)}{-q^2} \, .
\end{eqnarray}
This relation allows to express structure functions in terms of the Fourier transform of a bilocal matrix element. 

\subsection{Traditional Light-Cone approach}
\label{sec:LCA}

If we take 
\begin{eqnarray}
x \equiv (z,0,0,z)\, ,
\end{eqnarray}
and the proton at rest, we have from eqs.~(\ref{bilocstru1}) and~(\ref{bilocstru2}) the spectral decomposition
\begin{eqnarray}
&&f (\lambda) =  \frac {m} {2 \pi} \sum_n |\langle n | \phi(0) |m\rangle|^2 \int_{- \infty}^{+ \infty} d z e^{i z (-m+E_n- {p_n}_z+ \lambda m)} =\nonumber \\
&&= \sum_n |\langle n | \phi(0) |m\rangle|^2 \delta \Big{(}\frac {E_n- {p_n}_z}{m}-1+ \lambda\Big{)} \label{spectrum} \, ,
\end{eqnarray}
where $|m\rangle$ is the state of a proton at rest.

\subsection{Equal-Time approach}
\label{sec:ETA}

In refs.~\cite{Ji:2013dva} it is proposed that the structure function may be computed from eq.~(\ref{proposal}). Introducing intermediate states, we get
\begin{eqnarray}
&&F (\omega) = \lim_{P_z \rightarrow +\infty} \frac {P_z}{2 \pi} \sum_n | \langle n| \phi(0)|P_z \rangle|^2 \int_{-\infty}^{+\infty} d z e^{iz \omega P_z} e^{-i z ({p_n}_z - P_z)} = \nonumber \\
&&= \lim_{P_z \rightarrow +\infty} \sum_n |\langle n| \phi(0)|P_z \rangle|^2 \delta \Big{(}\omega +1- \frac {{p_n}_z}{P_z}\Big{)} \, .\label{FORET}
\end{eqnarray}
To make contact with the expression~(\ref{spectrum}) it is convenient to transfer the Lorentz transformation from the proton to the space-time arguments of operators. Using eqs.~(\ref{LOR}) and~(\ref{BLOR}), we find
\begin{eqnarray}
&& \langle P_z|\phi (0) \phi(z)|P_z \rangle = \langle m| \phi (0) \phi (\frac {P_z} {m} z, 0, 0, \frac {\sqrt {m^2 +P_z^2}}{m} z )|m \rangle = \nonumber \\
&&= \sum_n |\langle n|\phi(0)|m \rangle|^2 e^{i (E_n - m)  \frac {P_z} {m} z} e^{-i p_{n_z} \frac {\sqrt {m^2 +P_z^2}}{m} z} \, .\label{SPECT}
\end{eqnarray}
From the definition~(\ref {proposal}) we therefore have 
\begin{eqnarray}
\hspace{-.8cm}&&F (\omega) \!= \! \!\lim_{P_z \rightarrow +\infty} \!P_z \sum_n |\langle n|\phi(0)|m \rangle|^2 \delta \Big{(}\omega P_z + (E_n - m)  \frac {P_z} {m}- {p_n}_z \frac {\sqrt {m^2 +P_z^2}}{m}\Big{)} \!= \nonumber \\
\hspace{-.8cm}&&= \!\lim_{P_z \rightarrow +\infty} \sum_n |\langle n|\phi(0)|m \rangle |^2 \delta \Big{(} \omega + \frac {(E_n - m)} {m}- {p_n}_z \frac {\sqrt {1 + \frac {m^2} {P_z^2}}}{m}\Big{)}\! = \nonumber \\
\hspace{-.8cm}&&=\! \sum_n |\langle n|\phi(0)|m\rangle|^2 \delta \Big{(}\omega - 1+ \frac {E_n - {{p_n}_z}} {m}\Big{)}\, , \label{ZFOR}
\end{eqnarray}
which indeed coincides with eq.~(\ref{spectrum}). This means that, barring renormalization effects, formula~(\ref{proposal}) correctly provides the Euclidean version of eq.~(\ref{spectrum}).

\appendix 
\renewcommand{\thesection}{Appendix B} 
\section{Trace operators in a toy-model}
\label{sec:APPB} 

To provide an intuition of the harm that the power divergent mixings can cause in the construction of parton distribution functions, we discuss a simple mathematical example mimicking what happens if divergent trace operators are not properly subtracted out in the process of renormalizing the leading twist DIS operators 
\begin{eqnarray}
O^{(n)}_{\mu_1 \dots \mu_n} = \phi (0) \partial_{\mu_1} \dots \partial_{\mu_n} \phi (0) \, .\label{DISOP}
\end{eqnarray}

1) If divergent mixings due to trace operators were absent, the $z$ dependence of the renormalized (finite) matrix elements of the $\phi(z)\phi(0)$ bilocal will only occur through the combination $P_z z$. With reference to the regularized theory after introducing the formal definition
\beq
G(P_z z;\Lambda) = \langle P_z|\phi(z)\phi(0)|P_z\rangle\Big{|}_\Lambda=\sum_{n}A^{(n)}(\Lambda)(P_z z)^n\, ,
\label{DEFI}
\eeq
the (regularized) parton distribution function will be given by
\beq
f(\omega;\Lambda) =P_z\int_{-\infty}^{\infty}dz\,e^{i\omega P_z z}G(P_z z,\Lambda) = 2\pi \sum_n (-1)^n A^{(n)}(\Lambda) \delta^{(n)}(\omega)\, ,
\label{FOMEGA}
\eeq
where 
\beq
\delta^{(n)}(\omega)=\frac{(-i)^n}{2\pi}\int_{-\infty}^{\infty} dx\, e^{i\omega x} x^n\, .
\label{DELTN}
\eeq
A few observations are in order here. First of all, we notice that eq.~(\ref{FOMEGA}) does not depend on $P$. Secondly, to make $f(\omega;\Lambda)$ a renormalization group invariant quantity one must proceed as described in sect.~\ref{sec:DR}, making use of the Mellin transform method to give to each moment its correct running.

2) In the presence of trace terms one needs to be more careful and explicit about   regularization. Thus we write for the matrix element, $G$, of the bilocal operator the integral representation
\beq
G(P_z z, z;\Lambda)=\int dk\, e^{-\frac{k^2}{\Lambda^2}}e^{ikz}g(P_z z,k) \, ,
\label{GREP}
\eeq
where the exponential factor $e^{ikz}$ has been introduced to describe the effects of trace operators. In fact, if Taylor-expanded, it gives rise to power divergent terms of the kind $(\Lambda z)^n$. 

In this toy-model the matrix element of the properly subtracted leading twist operator is then obtained by just crossing out the $e^{ikz}$ factor from the the previous equation. If we do so, eq.~(\ref{GREP}) leads to the parton distribution function 
\beq
f(\omega;\Lambda) =P_z\int_{-\infty}^{\infty}dz\,e^{i\omega P_z z}  \int dk\, e^{-\frac{k^2}{\Lambda^2}}g(P_z z,k)=\int dk\, e^{-\frac{k^2}{\Lambda^2}} \,\tilde g(\omega,k) \, ,
\label{FOMEGAKT}
\eeq
where we have introduced the Fourier transform of $g$ with respect to its first argument
\beq
\tilde g(\omega,k)=\int dy \,e^{i\omega y} g(y,k)\, .
\label{FFG}
\eeq
Viceversa eq.~(\ref{GREP}) leads to the parton distribution function 
\beqn
&&\widehat f(\omega;\Lambda) =P_z\int_{-\infty}^{\infty}dz\,e^{i\omega P_z z}  \int dk\, e^{-\frac{k^2}{\Lambda^2}}e^{ikz}g(P_z z,k) =\nonumber\\
&&=\int dk\, e^{-\frac{k^2}{\Lambda^2}} \tilde g(\omega+\frac{k}{P_z},k) \, .
\label{FOMEGAK}
\eeqn
Owing to the Riemann--Lebesgue lemma the $k$-integral in eq.~(\ref{FOMEGAK}) converges even in the limit $\Lambda\to\infty$. This analysis proves that mixings with trace operators do not show up as (power) divergences in the structure function (see eq.~(\ref{FOMEGAK})). Rather at finite $P_z$ they deform the expression of the latter. Unfortunately in the lattice regularization one cannot send $P_z$ to infinity as $P_z$ can never be made larger than $a^{-1}$.

3) Within the simple toy-model we are discussing it is not difficult to see that, if DIS operators are made finite with the proper subtractions of power divergent trace operators, the remaining finite trace operator contributions to the structure function do indeed vanish in the limit of large $P_z$. 

In fact, the situation in which power divergent trace operator mixings are subtracted out from the bare DIS operator can be mimicked by stipulating that the function $g(P_z z,k)$ has a well convergent behaviour for large $k$ with an exponential cutoff scaled by some physical, finite mass parameter, $\Lambda_s$. Thus, assuming for $g(P_z z,k)$ the behaviour~\footnote{For the sake of this argument one might equivalently well assume $g(P_z z,k)\sim e^{-\frac{k^2}{\Lambda_s^2}} h (P_z z,k)$.}
\beq
g(P_z z,k)\sim e^{-\frac{|k|}{\Lambda_s}} h(P_z z,k)
\label{GTO}
\eeq
with $h(P_z z,k)$ a smooth, bounded function of $k$, we can immediately send the u.v.\ cutoff, $\Lambda$ to infinity as the $k$ integral is convergent. In this situation one gets 
\beqn
\hspace{-.8cm}&&P_z\int_{-\infty}^{\infty}dz\,e^{i\omega P_z z}  \int dk\, e^{ikz} e^{-\frac{|k|}{\Lambda_s}}h(P_z z,k) \stackrel{P_z\gg\Lambda_s}\longrightarrow\nonumber\\
\hspace{-.8cm}&&\stackrel{P_z\gg\Lambda_s}\longrightarrow\int dk\, e^{-\frac{|k|}{\Lambda_s}} \tilde h(\omega+\frac{k}{P_z},k)=\int dk\,\tilde g(\omega,k)\, .
\label{FOMEGATR}
\eeqn
The last expression exactly coincides with the last equality in eq.~(\ref{FOMEGAKT}) after removing the u.v.\ cutoff. We recall that we can safely take the limit $\Lambda\to\infty$ in eq.~(\ref{FOMEGAKT}) as the latter represents the expression of the structure function in the case trace operator mixings are absent.


\thebibliography{99}

\bibitem{Ji:2013dva}
  X.~Ji,
  Phys.\ Rev.\ Lett.\  {\bf 110} (2013) 262002.

\bibitem{Yndurain}
F.~J.~Yndur\`ain, ``The Theory of Quark and Gluon Interactions'', Theoretical and Mathematical Physics, ISBN-13 978-3-540-33209-1 4th Edition (Springer Berlin Heidelberg New York, 2006).

\bibitem{Xiong:2013bka} 
  X.~Xiong, X.~Ji, J.~H.~Zhang and Y.~Zhao,
  Phys.\ Rev.\ D {\bf 90}, 014051 (2014).

\bibitem{Ma:2014jga} 
  Y.~Q.~Ma and J.~W.~Qiu,
  Int.\ J.\ Mod.\ Phys.\ Conf.\ Ser.\  {\bf 37}, 1560041 (2015).

\bibitem{Lin:2014zya}
  H.~W.~Lin, J.~W.~Chen, S.~D.~Cohen and X.~Ji,
  Phys.\ Rev.\ D {\bf 91} (2015) 054510.

\bibitem{Ji:2015jwa} 
  X.~Ji and J.~H.~Zhang,
  Phys.\ Rev.\ D {\bf 92}, 034006 (2015).

\bibitem{Alexandrou:2015rja}
  C.~Alexandrou, K.~Cichy, V.~Drach, E.~Garcia-Ramos, K.~Hadjiyiannakou, K.~Jansen, F.~Steffens and C.~Wiese,
  Phys.\ Rev.\ D {\bf 92} (2015) 014502.

\bibitem{Zhang:2017bzy}
  J.~H.~Zhang, J.~W.~Chen, X.~Ji, L.~Jin and H.~W.~Lin,
  Phys.\ Rev.\ D {\bf 95} (2017) no.9,  094514.
  
\bibitem{Alexandrou:2017huk}
  C.~Alexandrou, K.~Cichy, M.~Constantinou, K.~Hadjiyiannakou, K.~Jansen, H.~Panagopoulos and F.~Steffens,
  arXiv:1706.00265 [hep-lat].

\bibitem{Orginos:2017kos}
  K.~Orginos, A.~Radyushkin, J.~Karpie and S.~Zafeiropoulos,
  arXiv:1706.05373 [hep-ph].
  
\bibitem{Brandt:1972nw}
  R.~A.~Brandt and G.~Preparata,
  Fortsch.\ Phys.\  {\bf 20} (1972) 571.

\bibitem{Fritzsch:2003fg}
  H.~Fritzsch and M.~Gell-Mann,
  hep-ph/0301127.

\bibitem{Christy:2012tk}
  M.~E.~Christy, J.~Blumlein and H.~Bottcher,
  arXiv:1201.0576 [hep-ph].

\bibitem{Capitani:1994qn}
  S.~Capitani and G.~Rossi,
  Nucl.\ Phys.\ B {\bf 433} (1995) 351.

\bibitem{Beccarini:1995iv}
  G.~Beccarini, M.~Bianchi, S.~Capitani and G.~Rossi,
  Nucl.\ Phys.\ B {\bf 456} (1995) 271.
  
  \bibitem{LIGHTHILL}
M.J. Lighthill, ``An introduction to Fourier analysis and generalised functions'', Cambridge Monographs on Mechanics and Applied Mathematics (Cambridge University Press, 1964).
  
\bibitem{Boucaud:1989ga}
  P.~Boucaud, L.~C.~Lung and O.~Pene,
  Phys.\ Rev.\ D {\bf 40} (1989) 1529.
  
\bibitem{Boucaud:1992nf}
  P.~Boucaud, J.~P.~Leroy, J.~Micheli, O.~P\`{e}ne and G.~C.~Rossi,
  Phys.\ Rev.\ D {\bf 47} (1993) 1206.
  
\bibitem{Martinelli:1995vj}
  G.~Martinelli and C.~T.~Sachrajda,
  Phys.\ Lett.\ B {\bf 354} (1995) 423.

\bibitem{Ishikawa:2016znu}
  T.~Ishikawa, Y.~Q.~Ma, J.~W.~Qiu and S.~Yoshida,
  arXiv:1609.02018 [hep-lat].
  
\bibitem{Ishikawa:2017jtf}
  T.~Ishikawa, Y.~Q.~Ma, J.~W.~Qiu and S.~Yoshida,
  PoS LATTICE {\bf 2016} (2016) 177
  [arXiv:1703.08699 [hep-lat]].
  
\bibitem{Chen:2016fxx}
  J.~W.~Chen, X.~Ji and J.~H.~Zhang,
  Nucl.\ Phys.\ B {\bf 915} (2017) 1.
\end{document}